\documentclass[12pt]{iopart}

\newcommand{\be}{\begin{eqnarray}}
\newcommand{\ee}{\end{eqnarray}}
\newcommand{\bea}{\begin{eqnarray}}
\newcommand{\eea}{\end{eqnarray}}

\usepackage{graphicx}
\usepackage{bm}
\usepackage{color}
\usepackage{slashed}
\usepackage{cite}
\usepackage{graphicx}
\usepackage{multicol}
\usepackage{graphicx}
\usepackage{color}
\usepackage{slashed}
\usepackage{CJKutf8}
\begin{document}
\begin{CJK}{UTF8}{<font>}
\title{A Novel Jet Model for the Novikov-Thorne Disk and its Observable Impact}

\author{Sen Guo$^{1}$, \ Pei Wang$^{1}$, \ Ke-Jian He$^{2}$, \ Guo-Ping Li$^{3}$, \ Xiao-Xiong Zeng$^{1,2}$,\ Wen-Hao Deng$^{3}$}

\address{$^1$College of Physics and Electronic Engineering, Chongqing Normal University, Chongqing 401331, People's Republic of China}
\address{$^2$School of Civil Engineering, Chongqing Jiaotong University, Chongqing 400074, People's Republic of China}
\address{$^3$School of Physics and Astronomy, China West Normal University, Nanchong 637000, People's Republic of China}

\ead{sguophys@126.com; wangpei19717@163.com; kjhe94@163.com; gpliphys@yeah.net; xxzengphysics@163.com;15760064548@163.com}
\vspace{10pt}
\begin{indented}
\item[]Mar. 2025
\end{indented}

\begin{abstract}
Recent high-resolution observations have established a strong link between black hole jets and accretion disk structures, particularly in the 3.5 mm wavelength band [Nature. 616, 686 (2023)]. In this work, we propose a ``jet-modified Novikov-Thorne disk model'' that explicitly incorporates jet luminosity into the accretion disk radiation framework. By integrating synchrotron radiation from relativistic electrons in the jet, we derive a modified luminosity function that accounts for both the accretion disk and jet contributions. Our analysis demonstrates that the inclusion of jet luminosity enhances the total accretion disk luminosity by approximately 33.5\%, as derived from the integration of radiative flux. Furthermore, we compare our modified model with the standard Novikov-Thorne model and find that the jet contribution remains significant across different observational inclinations. These results highlight the necessity of incorporating jet effects when estimating the observable flux of black hole accretion systems, which has direct implications for future astronomical observations.
\end{abstract}


\section{Introduction}
\label{intro}
\par
The study of black holes has advanced significantly in recent decades, primarily due to breakthroughs in observational astronomy and numerical simulations. The detection of gravitational waves from black hole mergers by LIGO/Virgo \cite{Abbott2016}, along with the direct imaging of event horizons by the Event Horizon Telescope (EHT), has provided compelling evidence for the existence of black holes and their strong-field relativistic effects \cite{Akiyama2019}.

\par
Astrophysical black holes are typically surrounded by accretion disks, which serve as the primary sources of electromagnetic radiation. The structure and dynamics of these disks have been extensively studied through numerical simulations, particularly in the context of magnetized accretion flows \cite{McKinney2012}. Moreover, high-resolution observations of M87* have revealed a bright, ring-like structure associated with the accretion flow and jet base, reinforcing the theoretical connection between jet dynamics and accretion disk physics \cite{Lu2023}.

\par
In 1963, Maarten Schmidt observed the spectrum of 3C-273 using the Palomar Telescope and discovered that its Balmer line redshift reached $z = 0.158$. This finding overturned the ``nearby star'' hypothesis and confirmed that quasars are extremely high-energy celestial objects on cosmological scales. With luminosities far exceeding those of normal galaxies, quasars prompted scholars to propose the supermassive black hole (SMBH) accretion hypothesis \cite{Schmidt1963}. Subsequently, Donald Lynden-Bell pioneered the concept of accretion disks, demonstrating how black holes release gravitational energy by accreting matter \cite{Lynden-Bell1969}. While Lynden-Bell＊s ideas established the foundation, the standard Shakura-Sunyaev Disk (SSD) model was mathematically formalized in 1973 by Shakura and Sunyaev \cite{Shakura1973}. In the same year, Novikov and Thorne enriched the SSD framework by incorporating general relativity, most critically through the concept of the innermost stable circular orbit (ISCO). This refinement clarified the disk＊s dynamical boundaries and energy extraction mechanisms \cite{Novikov1973}. Soon after, Thorne and Page systematically introduced general relativity into accretion disk theory for the first time, laying the cornerstone of modern black hole accretion disk models. They re-derived the thin-disk structure within the relativistic framework, established rigorous mathematical formulations for mass and angular momentum transport as well as radiative efficiency, paving the way for the practical application of the standard accretion disk model (SSD) \cite{Page1974}. Then Shapiro proposed the two-temperature accretion disk model, resolving the high-energy spectrum puzzle of black hole X-ray binaries. By abandoning the standard thin-disk thermal equilibrium assumption, they addressed the low-luminosity black hole spectra unexplained by standard thin disks and successfully predicted the origin of hard X-ray radiation and radio jets \cite{Shapiro1976}.

\par
The Blandford-Znajek (BZ) mechanism, proposed in 1977, became the cornerstone of relativistic jet physics by leveraging black hole spin-energy extraction via magnetic fields \cite{Blandford1977}. In the next year, Abramowicz pioneered the mathematical model of thick accretion disks, revealing novel accretion regimes under extreme conditions through rigorous hydrodynamic analysis. His proposed toroidal structures and sub-Keplerian rotation concepts remain foundational tools in high-energy astrophysics, providing irreplaceable theoretical value for understanding super-Eddington accretion, thermal accretion flows, and multi-wavelength radiation mechanisms \cite{Abramowicz1978}.
Blandford and Konigl then expanded this paradigm by constructing a relativistic jet analytical model. Assuming a power-law electron energy distribution ($N(\gamma) \propto \gamma^{-p}$), they derived the relationship between radio spectral indices and the electron index $p$, laying the groundwork for unified jet radiation theory \cite{Blandford1979}. 
Later, in 1982, Blandford and Payne introduced the Blandford-Payne (BP) mechanism, explaining centrifugal launching of jets from accretion disks. This complemented the BZ mechanism by addressing jets in non- or weakly spinning black hole systems \cite{Blandford1982}.

\par
In 1985, Uchida and Shibata were the first to use numerical simulations to demonstrate how magnetohydrodynamic (MHD) processes generate collimated jets from accretion disks. This work provided the first numerical verification of the magnetic field-driven jet hypothesis and has inspired decades of subsequent research on jet dynamics \cite{Uchida1985}. 
Stone and Norman then developed the ZEUS-2D numerical code, followed later by the implementation of ZEUS-3D. Utilizing this 3D MHD framework, Stone and Hardee simulated magnetized astrophysical jets and discovered that their propagation is significantly disrupted by kink-mode instabilities, leading to lateral distortion and fragmentation. These simulations predicted the formation of knot-like structures closely resembling the observed quasi-periodic shock features and bright knots in radio galaxies such as M87 \cite{Stone2000}. 
By 1994, Rees＊ fireball model tied jet dynamics to ultrarelativistic expansion, explaining gamma-ray burst (GRB) prompt emission and later evolving into baryonic jet models to account for heavy particle-dominated outflows \cite{Rees1994}. At the same time, Narayan and Yi systematically developed the first self-similar analytical model for advection-dominated accretion flows (ADAF), revolutionizing the theoretical framework for low-accretion-rate systems. They unified viscous heating, electron-ion energy exchange, and radial energy advection under a two-temperature framework for the first time, extending the Shakura每Sunyaev $\alpha$-viscosity hypothesis to ADAF \cite{Narayan1994}.

\par
In 1999, Koide pioneered the implementation of general relativistic magnetohydrodynamics (GRMHD) into numerical simulations of black hole jets. Building upon this foundation, Gammie, McKinney, and collaborators introduced the HARM (High-Accuracy Relativistic Magnetohydrodynamics) code in 2003 - the first fully general relativistic MHD framework capable of self-consistently modeling black hole accretion flows and their coupled jets within curved spacetime, thereby validating the governing equations of relativistic plasma dynamics in strong gravitational fields \cite{Gammie2003}. Not long after, Komissarov proposed an analytical solution model for magnetized tori around black holes, unveiling the profound impact of magnetic fields on accretion disk morphology and stability \cite{Komissarov2006}.
Subsequently, Tchekhovskoy, Narayan, and McKinney leveraged HARM to systematically simulate black holes across a range of spin parameters. Their computations revealed a striking dependence of normalized jet efficiency on spin: for extreme spins ($a = 0.99$), the BZ mechanism dominates, surpassing the canonical $10\%$ accretion energy limit; conversely, for non-spinning holes ($a = 0$), jet power originates solely from disk magnetic dissipation. These results align with statistical trends in radio-loud quasars, where high-spin systems exhibit significantly higher jet power at fixed accretion rates \cite{Tchekhovskoy2011}. 
In 2012, McKinney＊s HARM simulations uncovered that under magnetically arrested disk (MAD) conditions, accumulated magnetic flux near the horizon saturates into flux-tube structures. The large-scale field anchored to these tubes continuously extracts rotational energy via the BZ process, explaining why rapidly spinning black holes launch ultra-relativistic, tightly collimated jets \cite{McKinney2012}. In the same year, Tejeda developed an innovative analytical framework to construct a concise relativistic accretion model for Kerr black holes, bridging the gap between complex numerical simulations and classical theories \cite{Tejeda2012}.

\par
Thereafter, Compere proposed a self-similar theory for steady-state thin accretion disks around near-extremal Kerr black holes. By combining the symmetries of extremal black hole spacetimes with accretion flow dynamics, he revealed the unique properties of accretion structures in such strong gravitational environments \cite{Compere2017}. 
Addressing computational limitations, Porth developed the BHAC code, incorporating adaptive mesh refinement (AMR) and multi-physics coupling. This advancement enabled high-fidelity studies of blazar jet morphology and black hole merger magnetospheric interactions \cite{Porth2017}. 
The 2019 EHT imaging of M87*＊s shadow, combined with polarized emission maps, resolved a toroidal magnetic field morphology at the jet base 〞 direct observational evidence corroborating BZ mechanism predictions \cite{Akiyama2019}. In a seminal follow-up investigation, Samik Mitra pioneered a breakthrough exploration of black hole accretion flows using general relativistic magnetohydrodynamic (GRMHD) simulations, employing high-fidelity numerical algorithms to unravel the nonlinear coupling between large-scale magnetic fields, extreme gravitational fields, and turbulent accretion plasmas \cite{Mitra2022}. A groundbreaking study recently presents an intriguing analytical formulation for steady-state axisymmetric magnetized accretion disk configurations surrounding Kerr black holes, transcending previous methodologies exclusively reliant on numerical simulations or weakly magnetized approximations \cite{Hou2023}.

\par
Several mechanisms have been proposed to explain jet formation and energy extraction. The BZ mechanism \cite{Blandford1977} describes how rotating black holes power jets via magnetic fields, whereas the BP mechanism \cite{Blandford1982} suggests that jets can originate from the accretion disk itself. Recent studies \cite{EHT2023} indicate that jet formation is strongly coupled to accretion flow properties, suggesting a non-negligible contribution of jet luminosity to the observed emission. However, most existing accretion disk models, including the classical Novikov-Thorne model \cite{Novikov1973, Page1974}, neglect the impact of jet radiation.

\par
To bridge this gap, we propose a ``jet-modified Novikov-Thorne disk model'' that explicitly incorporates jet luminosity into the accretion disk radiation framework. By integrating synchrotron emission from relativistic electrons in the jet \cite{Ginzburg1965}, we derive a modified luminosity function and analyze its impact on the observed flux. Our work aims to provide a more complete description of accretion disk radiation, which is crucial for accurately interpreting black hole observations.

\section{Accretion Disk Luminosity with Jet Contribution}
\label{sec:2}
\par
In this section, we investigate a static spherically symmetric black hole metric described by the equation:
\begin{equation}
\label{2-1}
{\rm d}s^{2}=-f(r){\rm d}t^{2}+f(r)^{-1}{\rm d}r^{2}+r^{2}({\rm d}\theta^{2}+\sin^{2}\theta {\rm d}\phi^{2}).
\end{equation}
Here, $f(r)$ represents the black hole metric potential, and it can be expressed as:
\begin{equation}
\label{2-2}
f(r) = 1 - \frac{2M}{r},
\end{equation}
where $M$ represents the mass of the black hole. Assuming a geometrically thin and optically thick accretion disk surrounding the black hole, we calculate the electromagnetic radiation flux ($\rm ergs^{-1}cm^{-2}str^{-1}Hz^{-1}$) emitted from a specific radial position $r$ on the disk using the following equation \cite{Luminet1979}:
\begin{equation}
\label{2-3}
F = - \frac{\dot{M}}{4\pi \sqrt{\rm -I}} \frac{\Omega_{,\rm r}}{(E-\Omega L)^{2}} \int_{r_{\rm in}}^{r} (E- \Omega L)L_{,\rm r} {\rm d} r,
\end{equation}
where $\dot{M}$ represents the mass accretion rate, $I$ is the determinant of the induced metric in the equatorial plane, and $r_{\rm in}$ denotes the inner edge of the accretion disk. The parameters $E$, $\Omega$, and $L$ represent the angular velocity, energy and angular momentum of the particles in a circular orbit, respectively, and they can be expressed as follows\cite{Bardeen1972}:
\begin{eqnarray}
\label{2-4}
&&E= - \frac{g_{\rm tt} + g_{\rm t \phi} \Omega}{\sqrt{-g_{\rm tt} + 2g_{\rm t \phi}\Omega - g_{\rm \phi \phi}\Omega^{2}}},\\
\label{2-5}
&&L= \frac{g_{\rm t \phi} + g_{\rm \phi \phi} \Omega}{\sqrt{-g_{\rm tt} + 2g_{\rm t \phi}\Omega - g_{\rm \phi \phi}\Omega^{2}}},\\
\label{2-6}
&&\Omega=\frac{{\rm d}\phi}{{\rm d}t}=\frac{-g'_{\rm t \phi} + \sqrt{(g'_{\rm t \phi})^{2} - g'_{\rm tt}g'_{\rm \phi \phi}}}{g'_{\rm \phi \phi}}.
\end{eqnarray}

\par
According to Eqs. (\ref{2-2})-(\ref{2-6}), the radiation flux over the disk for Schwarzschild black hole is \cite{Page1974}
\begin{equation}
\label{2-7}
F_{\rm disk}(r) = - \frac{3 \dot{M} \sqrt{{M}/{r^{3}}}}{\pi r(24M - 8r)} \int_{r_{\rm in}}^{r} \frac{(6M-r)\sqrt{{M}/{r^{3}}}r}{6M-2r} dr.
\end{equation}

\par
On the other hand, the luminosity of a black hole jet is produced by high-energy particles that emit synchrotron radiation in a magnetic field. The resulting radiation spectrum depends on factors such as the magnetic field strength, the energy distribution of electrons, and the angle between the electron velocity and the magnetic field. The total luminosity of the jet is obtained by integrating the synchrotron radiation from all electrons in the jet. By combining the accretion disk and jet luminosity models, we investigate how the jet affects the optical appearance of the accretion disk. To construct our jet model, we assume that the relativistic electrons in the jet follow a power-law energy distribution \cite{Blandford1979}:
\begin{equation}
N(\gamma) = N_0 \gamma^{-p},
\end{equation}
where $\gamma$ is the Lorentz factor, $p$ is the spectral index, and $N_0$ is the normalization factor. The synchrotron radiation power emitted per unit frequency is given by \cite{Rybicki1979}:
\begin{equation}
P_{\nu} = \frac{\sqrt{3} q_e^3 B \sin{\theta}}{m_e c^2}
\int\limits_{\gamma_{\min}}^{\gamma_{\max}} N(\gamma)
F\left( \frac{\nu}{\nu_c(\gamma)} \right) d\gamma,
\end{equation}
where $B$ is the magnetic field strength, $\theta$ is the pitch angle, and $\nu_c$ is the critical frequency\cite{Rybicki1979}.
\begin{equation}
\nu_c(\gamma) = \frac{3 q_e B}{4 \pi m_e c} \gamma^2.
\end{equation}
Full derivations are provided in Appendix A. We incorporate this emission mechanism into the Novikov-Thorne model to account for the additional contribution of the jet to the total flux. Furthermore, the interaction between the jet and the accretion disk introduces an additional radiative component due to Compton upscattering of disk photons by relativistic electrons in the jet. This process enhances the observed flux and modifies the spectral energy distribution (SED) of the accretion disk emission. The observed flux incorporating jet contributions is given by:
\begin{equation}
F_{\rm total} = F_{\rm disk} + F_{\rm jet},
\end{equation}
where $F_{\rm jet}$ includes both synchrotron and inverse Compton contributions.

\begin{figure}[h]
  \centering
  \includegraphics[width=6cm,height=6cm]{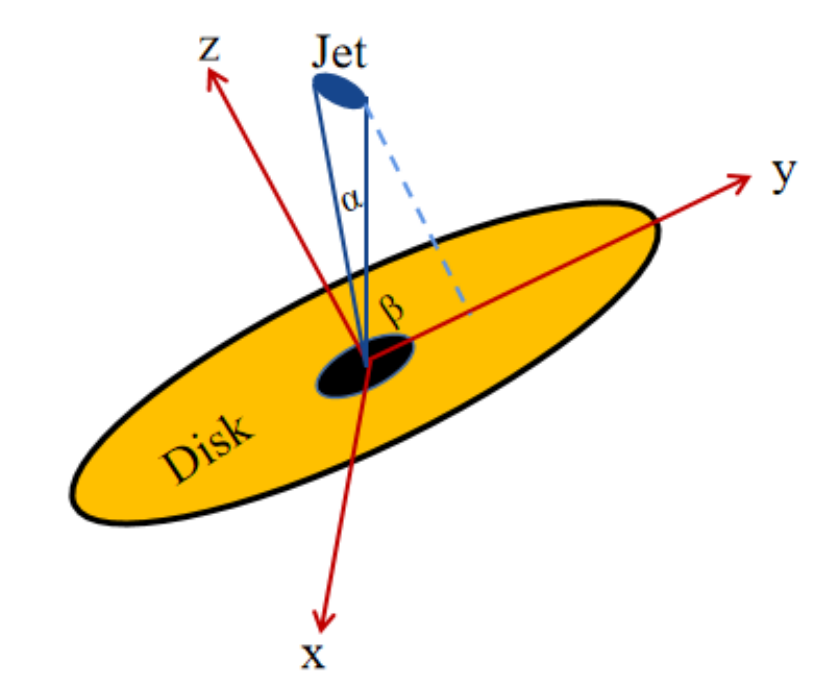}
  \caption{Jet model of Novikov-Thorne disk.}\label{fig:1}
\end{figure}

\par
\begin{figure*}[htbp]
  \centering
  \includegraphics[width=5cm,height=4cm]{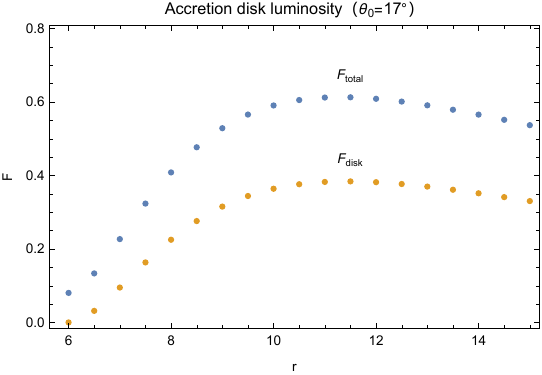}
  \hspace{0.1cm}
  \includegraphics[width=5cm,height=4cm]{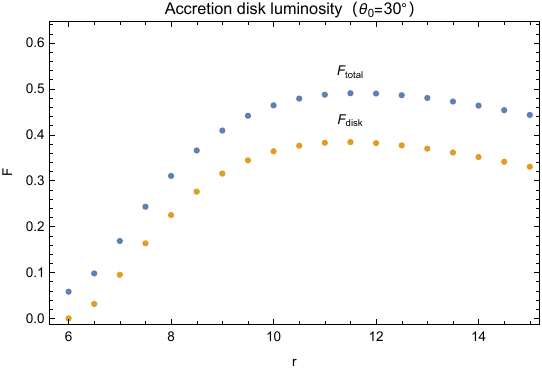}
  \hspace{0.1cm}
  \includegraphics[width=5cm,height=4cm]{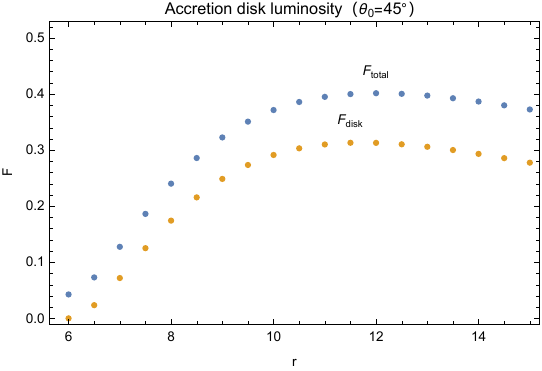}
  \caption{Observable luminosity of the Novikov-Thorne model and jet model of Novikov-Thorne disk.}\label{fig:2}
\end{figure*}

To visualize the effects of jet contributions, we compare the standard Novikov-Thorne model with our modified model in Figure \ref{fig:2}. It is evident that jet radiation significantly enhances the observed luminosity, particularly at higher frequencies.

\section{Accretion Disk Image}
\label{sec:3}
To obtain the observable radiant flux for a distant observer, the correction for gravitational redshift must be considered. The redshift factor, denoted by $d = 1 + z$, is the ratio of the emitted photon energy $E_{\rm em}$ to the observed photon energy $E_{\rm obs}$ and accounts for the gravitational potential difference between the source and observer. In general relativity, the projection of photon four-momentum $k_{\mu}$ onto the four-velocities of the source $p_{\rm source}^{\mu}$ and observer $p_{\rm obs}^{\mu}$ must be taken into account to calculate $E_{\rm obs}$. The redshift factor can be calculated using the following formula \cite{14}:
\begin{equation}
\label{2-10}
d = 1 + z = \frac{E_{\rm em}}{E_{\rm obs}}.
\end{equation}
The emitted energy $E_{\rm em}$ is given by
\begin{equation}
\label{2-11}
E_{\rm em} = p_{\rm t} u^{\rm t} + p_{\rm \phi} u^{\rm \phi} = p_{\rm t} u^{\rm t} \Bigg(1 + \Omega \frac{p_{\rm \phi}}{p_{\rm t}}\Bigg),
\end{equation}
where $p_{\rm t}$ and $p_{\rm \phi}$ represent the photon four-momentum. For observers at a large distance, the ratio $p_{\rm t}/p_{\rm \phi}$ represents the impact parameter of the photons relative to the z-axis. The redshift factor $d$ can be rewritten as
\begin{equation}
\label{2-12}
d = 1 + z = \frac{E_{\rm em}}{E_{\rm obs}} = \frac{1 + b \Omega \cos \theta_{0}}{\sqrt{-g_{\rm tt} - 2 g_{\rm t \phi} g^{\rm t \phi} - g_{\rm \phi\phi}}},
\end{equation}
where $b$ represents the impact parameter, derived from the determination of the photon orbit around the black hole. $\theta_{0}$ is the observing inclination angle.

\par
From the above discussion, the observed flux of the accretion disk can be calculated using the redshift factor. The observed flux is expressed as \cite{Cunningham1975}
\begin{equation}
\label{2-13}
F_{\rm obs} = \frac{F_{\rm total}}{(1 + z)^{4}}.
\end{equation}

\par
Our model accounts for the contribution of the jet to the accretion disk luminosity, resulting in a brighter accretion disk image for a distant observer compared to the standard Novikov-Thorne model, which considers only accretion material. We present the impact of the jet flow on the observed brightness of the accretion disk, as shown in Figure 2. Our findings demonstrate noticeable differences in brightness at various observation angles ($\theta_{0}$). Specifically, we show that the jet flow contributes approximately $25.3\%$ of the total luminosity, resulting in a $33.8\%$ increase in the brightness of the accretion disk compared to the standard Novikov-Thorne model.

\par
We employed an optical tracking code to generate visual representations of a Novikov-Thorne disk model that includes jet contributions. Our results, as depicted in Figure 3, demonstrate a distinct concentration of brightness on the left side of the accretion disk, a consequence of the complex geometric interplay between the movement of matter and the propagation of light within the disk. Notably, there is also a pronounced accumulation of brightness in the northeast region of the image, which we attribute to the jet. To further investigate this phenomenon, we captured images from different observation angles and found that the additional luminosity persists regardless of the observation angle, confirming its dependence on the physical properties of the source itself. Furthermore, our analysis revealed a significant increase in brightness due to the jet flow in the southwest region of the secondary image, as shown in Figure 4.
\begin{figure}[htbp]
  \centering
  \includegraphics[width=9cm,height=8cm]{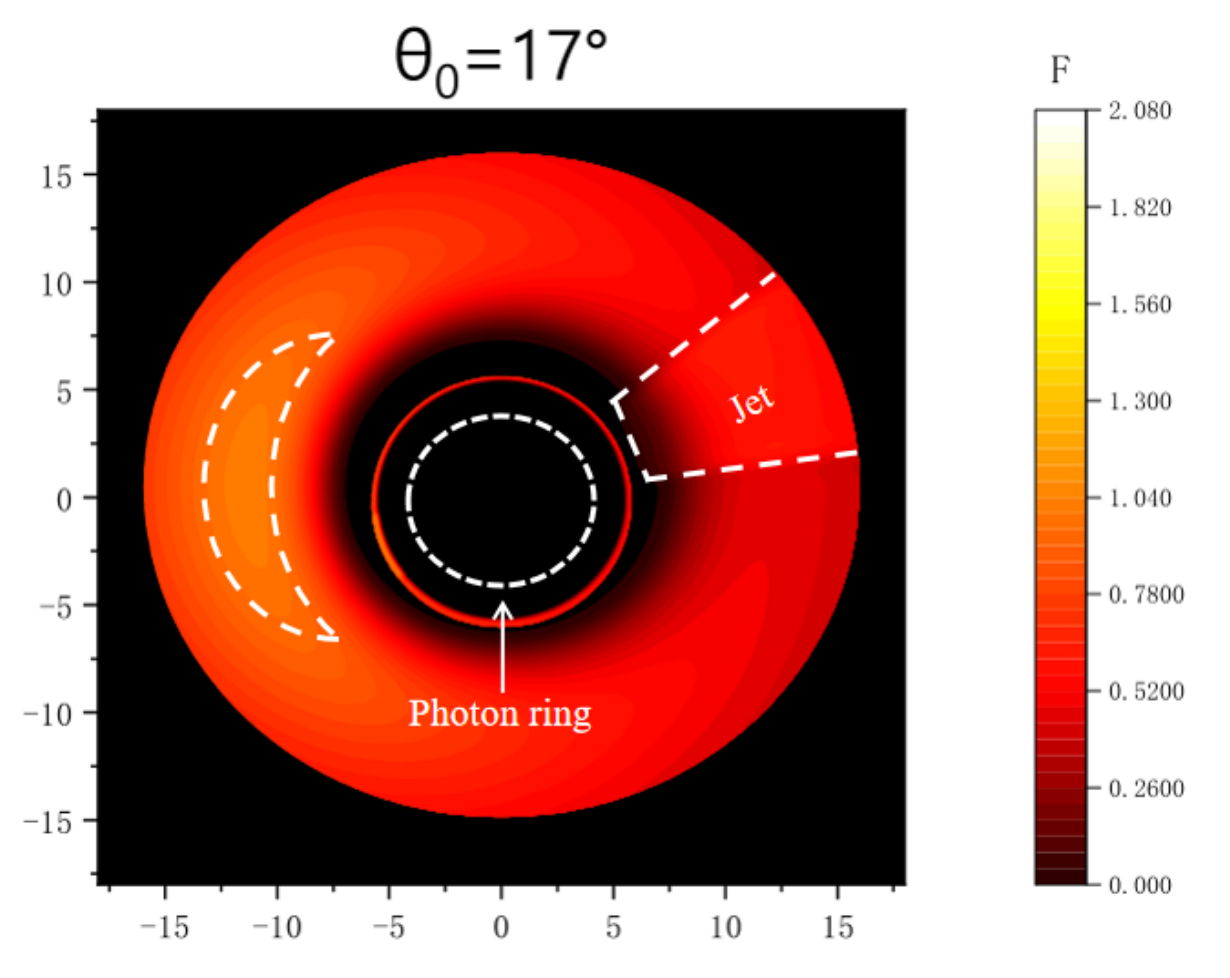}
  \caption {Accretion disk image including both disk and jet radiation.}\label{fig:3}
\end{figure}
\begin{figure*}[htbp]
  \centering
  \includegraphics[width=5cm,height=4.5cm]{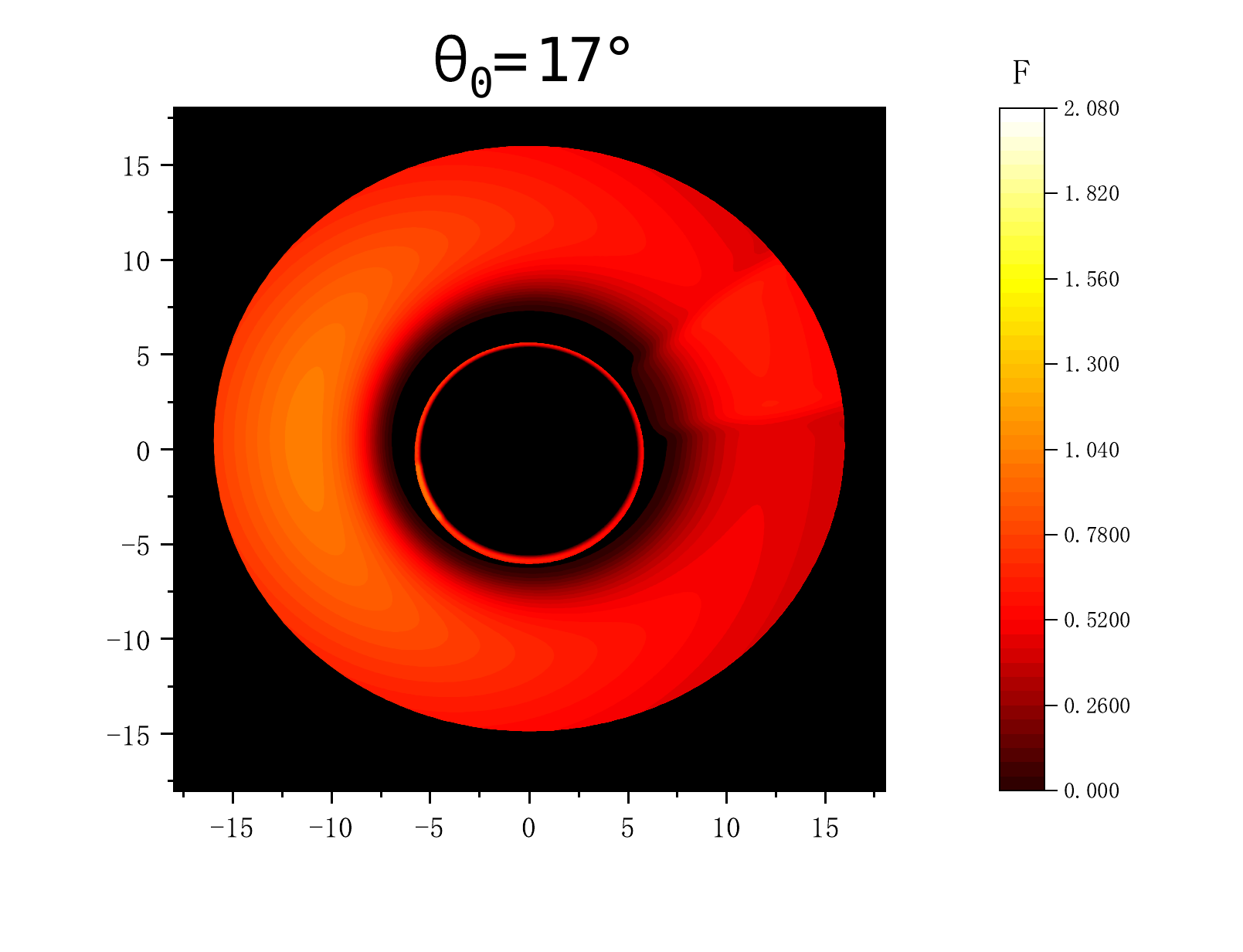}
  \includegraphics[width=5cm,height=4.5cm]{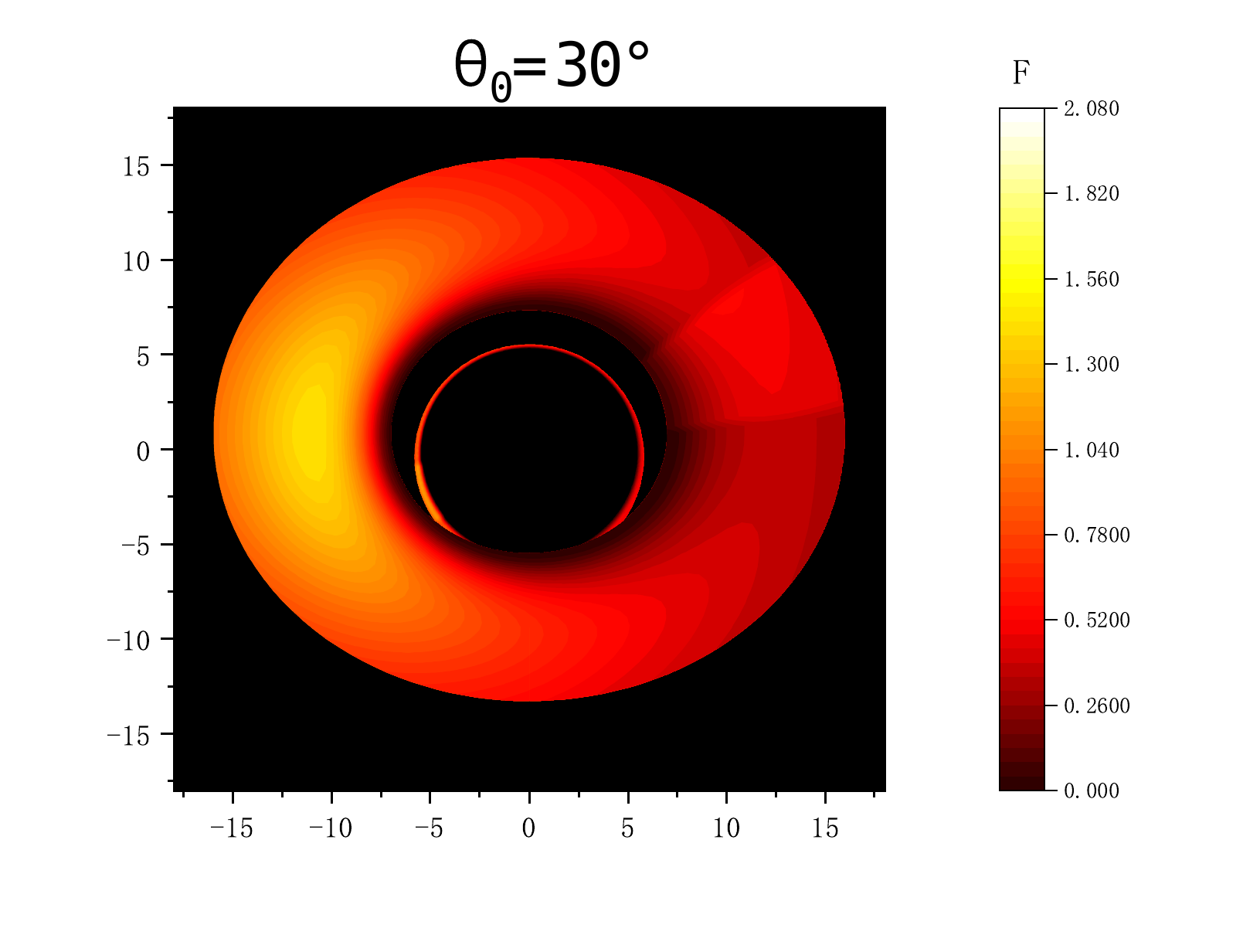}
  \includegraphics[width=5cm,height=4.5cm]{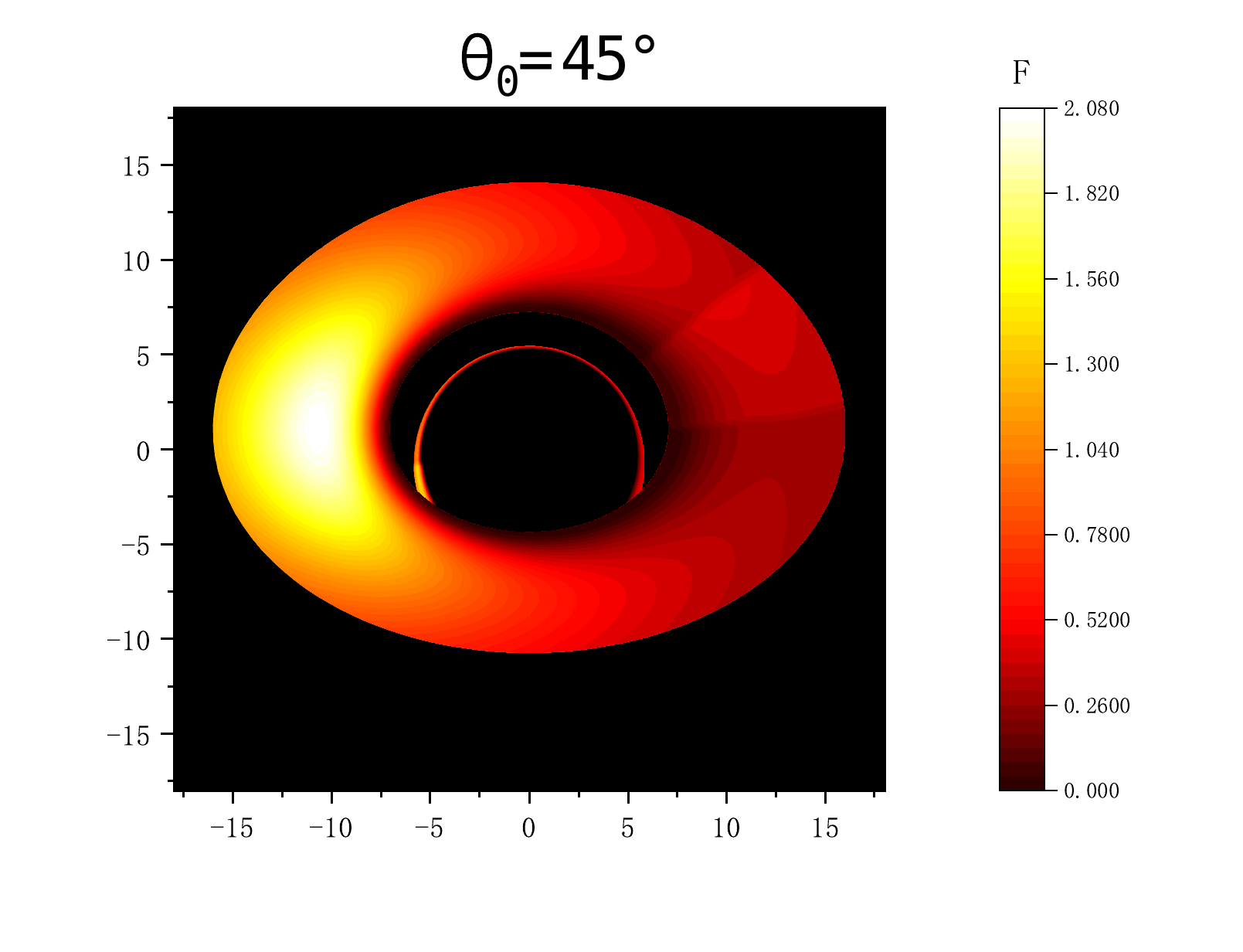}
  \caption {Images of an accretion disk with different observation angles. The position of the secondary image corresponds to the inner edge of the disk at $r_{\rm in} = 6M$, and the outer edge of the disk is at $r_{\rm out} = 15M$.}\label{fig:4}
\end{figure*}
%The BH mass is taken as $M = M_{\odot}$.
\begin{figure*}[htbp]
  \centering
  \includegraphics[width=4cm,height=4cm]{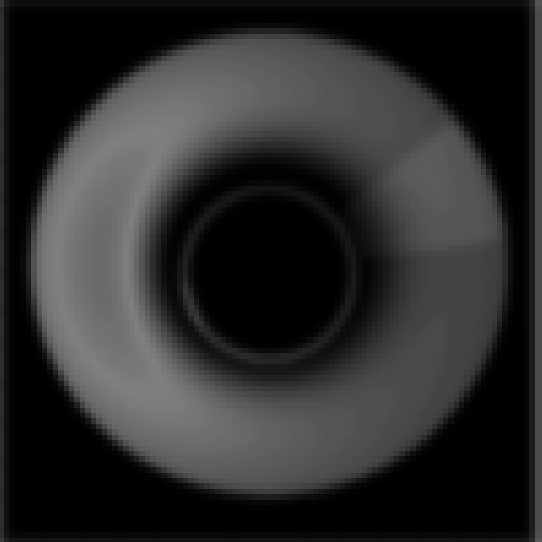}
  \hspace{1cm}
  \includegraphics[width=4cm,height=4cm]{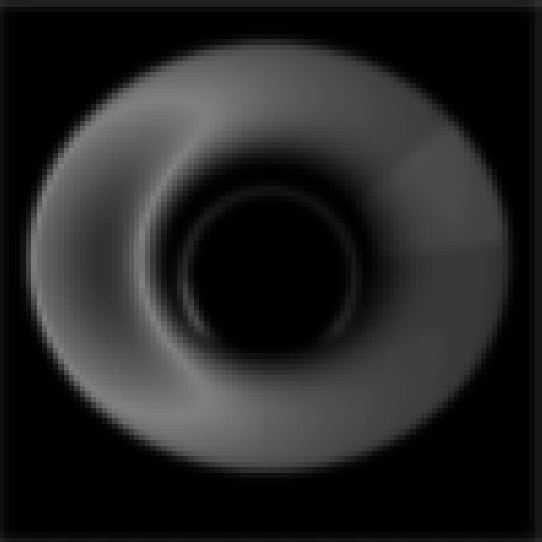}
  \hspace{1cm}
  \includegraphics[width=4cm,height=4cm]{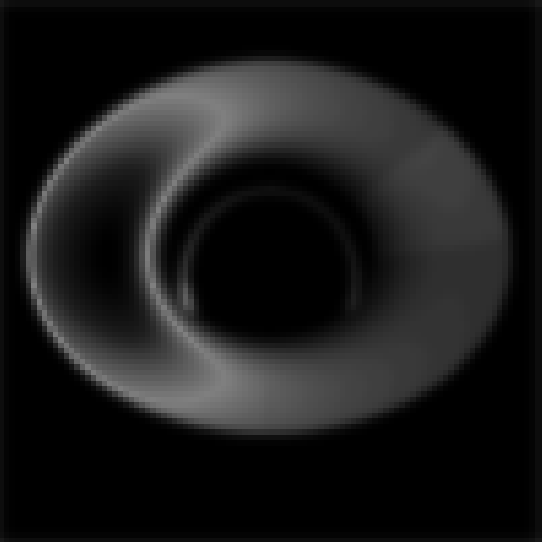}
  \caption {Blurred two-dimensional images processed using a Gaussian filter with a standard deviation of $1/12$ of the field of view.}\label{fig:5}
\end{figure*}

\par
We applied a Gaussian blur to the jet model of the Novikov-Thorne disk and obtained the image shown in Figure 5. This image clearly shows the black hole shadow as a dark region surrounding the black hole. The halo surrounding the shadow is also significant and results from the gravitational lensing effect. Moreover, our analysis identified two bright regions in the western and northeastern directions of the direct image, which can be attributed to the accretion and jet flow of matter around the black hole. This result also revealed that the observed images of the accretion disk are closely related to the observing inclination angle.

\section{Outlook}
\label{sec:4}
\par
In this letter, we introduce a novel jet model of the Novikov-Thorne disk and derive a general formula for the accretion disk luminosity that includes the jet contribution. Our findings demonstrate that black hole jets have a significant impact on the observable characteristics of black holes. While the motion of material in the accretion disk of a black hole is influenced by its gravitational and magnetic fields, resulting in the emission of radiation such as X-rays and ultraviolet light, the contribution of black hole jet luminosity to this radiation has not been taken into account in the conventional Novikov-Thorne model. Consequently, the observable flux function can be written as
\begin{equation}
F_{\rm total} = F_{\rm disk} + F_{\rm jet}.
\end{equation}

\par
Our analysis demonstrates that after incorporating the black hole jet luminosity, the luminosity of black hole radiation can increase by approximately $33.8\%$. This result suggests that accurately estimating the luminosity of black hole jets can provide a means of determining the observable radiant flux. The contribution of the jet persists at different observed inclinations, as shown by our comparison of the Novikov-Thorne model with our jet model. By extending the Novikov-Thorne model, our approach provides a more precise representation of observational data. Our work underscores the importance of considering black hole jet luminosity when estimating the observable radiant flux, which could advance our understanding of black hole behavior.

\par
Our findings indicate that the inclusion of jet luminosity leads to a substantial increase in the observed brightness of the accretion disk. This is consistent with recent EHT observations of M87*, which revealed that jet-disk interactions significantly enhance millimeter-wavelength emissions \cite{EHT2023}.

\par
Compared to previous models \cite{McKinney2012, Noble2009}, our approach explicitly accounts for the angle-dependent jet contribution to the observed flux. However, our model does not yet incorporate the time-dependent evolution of the jet, which can be explored in future work using full GRMHD simulations. Additionally, investigating the impact of jet precession and variability could further refine our understanding of jet-disk interactions.

\appendix 

\section{Equation derivation}

\par
Here we present the detailed derivation of Equations (9)-(10). For a single relativistic electron, the spectral power peaks near the critical frequency $\nu_c$, governed by the function:
\begin{equation}
P_{\nu}(\gamma)\propto\gamma^2B^2F \Big(\frac{\nu}{\nu_c(\gamma)}\Big),
\end{equation}
Maximizing $F(x)$ at $x = 1$ (i.e., $\nu = \nu_c$) gives:
\begin{equation}
\frac{dF(x)}{dx}\mid_{x=1}=0,
\end{equation}
to
\begin{equation}
 \nu=\nu_c(\gamma).
\end{equation}
Nonrelativistic gyrofrequency:
\begin{equation}
\nu_g=\frac{q_eB}{2\pi m_ec}.
\end{equation}
Time dilation increases the observed orbital period by $\gamma$:
\begin{equation}
\Delta t_{obs}=\gamma\Delta t_{rest}.
\end{equation}
Lorentz contraction shortens radiation pulse duration by $\gamma$:
\begin{equation}
\Delta t\sim\frac{1}{\gamma^2\nu_g}.
\end{equation}
Combined, the observed critical frequency scales as:
\begin{equation}
\nu_c(\gamma)\propto\gamma^2\nu_g=\gamma^2\frac{q_eB}{2\pi m_ec}.
\end{equation}
By rigorously computing the Fourier transform of the radiation field, the peak corresponds to the integral of the Bessel function $K_{5/3}(x)$:
\begin{equation}
F(\frac{\nu}{\nu_c})\propto\int_{0}^{\infty}K_\frac{5}{3}(x)dx.
\end{equation}
which introduces the factor $\frac{3}{4\pi}$. Substituting back:
\begin{equation}
\nu_c(\gamma) = \frac{3 q_e B}{4 \pi m_e c} \gamma^2.
\end{equation}

\section*{Acknowledgments}
This work is supported by the Fund Project of Chongqing Normal University (Grant Number: 24XLB033).

\clearpage

\end{CJK}

\section{References}
\addcontentsline{toc}{chapter}{References}
\begin{thebibliography}{99}\footnotesize
\itemsep=-3pt plus.2pt minus.2pt   % set the reference line spacing

% Black hole detection
\bibitem{Abbott2016}
B. P. Abbott et al. (LIGO Scientific Collaboration and Virgo Collaboration), ``Observation of Gravitational Waves from a Binary Black Hole Merger,'' \emph{Phys. Rev. Lett.}, vol. 116, no. 6, p. 061102, 2016.

\bibitem{Akiyama2019}
K. Akiyama et al. (Event Horizon Telescope Collaboration), ``First M87 Event Horizon Telescope Results. I. The Shadow of the Supermassive Black Hole,'' \emph{Astrophys. J. Lett.}, vol. 875, no. 1, p. L1, 2019.

\bibitem{McKinney2012}
J. C. McKinney, A. Tchekhovskoy, and R. D. Blandford, ``General relativistic magnetohydrodynamic simulations of magnetically choked accretion flows around black holes,'' \emph{Mon. Not. R. Astron. Soc.}, vol. 423, no. 4, pp. 3083每3117, 2012.

\bibitem{Lu2023}
R.-S. Lu et al., ``A Ring-like Accretion Structure in M87 Connecting its Black Hole and Jet,'' \emph{Nature}, vol. 616, pp. 686每690, 2023.

\bibitem{Schmidt1963}
Schmidt, M.,``3C 273 : A Star-Like Object with Large Red-Shift,''\emph{Nature}, vol. 197, pp. 1040, 1963.

\bibitem{Lynden-Bell1969}
Lynden-Bell, D.,``Galactic nuclei as collapsed old quasars,''\emph{Nature}, vol. 223, pp. 690, 1969.

\bibitem{Shakura1973}
Shakura, N. I. and Sunyaev, R. A.,``Black Holes in Binary Systems: Observational Appearances,''\emph{IAU Symp.}, vol. 55, pp. 155-164, 1973.

\bibitem{Novikov1973}
I. D. Novikov and K. S. Thorne, ``Astrophysics of black holes,'' in \emph{Black Holes}, C. DeWitt and B. S. DeWitt, Eds. Gordon and Breach, 1973.

\bibitem{Page1974}
D. N. Page and K. S. Thorne, ``Disk-Accretion onto a Black Hole. Time-Averaged Structure of Accretion Disk,'' \emph{Astrophys. J.}, vol. 191, pp. 499每506, 1974.

\bibitem{Shapiro1976}
Shapiro, S. L. and Lightman, A. P. and Eardley, D. M.,``A two - temperature accretion disk model for Cygnus X-1. 1. Structure and spectrum,''\emph{Astrophys. J.}, vol. 204, pp. 187-199, 1976.

\bibitem{Abramowicz1978}
Marek A. Abramowicz﹜Wojciech H. Zurek﹜Jean-Pierre Lasota,``Thick Accretion Disks with Purely Toroidal Flows,''\emph{Astronomy and Astrophysics}, vol. 63, pp. 221-224, 1978.

\bibitem{Blandford1977}
R. D. Blandford and R. L. Znajek, ``Electromagnetic Extraction of Energy from Kerr Black Holes,'' \emph{Mon. Not. R. Astron. Soc.}, vol. 179, no. 3, pp. 433每456, 1977.

\bibitem{Blandford1979}
Blandford, R. D. and Konigl, A.,``Relativistic jets as compact radio sources,''\emph{Astrophys. J.}, vol. 232, pp. 34, 1979.

\bibitem{Blandford1982}
R. D. Blandford and D. G. Payne, ``Hydromagnetic flows from accretion disks and the production of radio jets,'' \emph{Mon. Not. R. Astron. Soc.}, vol. 199, pp. 883每903, 1982.

\bibitem{Uchida1985}
Uchida, Y.,  Shibata, K.,``Magnetic jet formation from accretion disks: A two-dimensional MHD simulation. ''\emph{Publications of the Astronomical Society of Japan}, vol. 37, pp. 515每535, 1985.

\bibitem{Stone2000}
Stone, James M. and Hardee, Philip E.,``Mhd models of axisymmetric protostellar jets,''\emph{Astrophys. J.}, vol. 540, pp. 192, 2000.

\bibitem{Rees1994}
Rees, M. J. and Meszaros, P.,``Unsteady outflow models for cosmological gamma-ray bursts,''\emph{Astrophys. J. Lett.}, vol. 430, pp. L93--L96, 1994.

\bibitem{Narayan1994}
Narayan, Ramesh and Yi, In-su,``Advection dominated accretion: A Selfsimilar solution,''\emph{Astrophys. J. Lett.}, vol. 428, pp. L13, 1994.

\bibitem{Koide1999}
Koide, Shinji and Meier, David L. and Shibata, Kazunari and Kudoh, Takahiro,``General relativistic simulations of jet formation in a rapidly rotating black hole magnetosphere,''\emph{Astrophys. J.}, vol. 536, pp. 668, 2000.

\bibitem{Gammie2003}
Gammie, Charles F. and McKinney, Jonathan C. and Toth, Gabor,``HARM: A Numerical scheme for general relativistic magnetohydrodynamics,''\emph{Astrophys. J.}, vol. 589, pp. 444-457, 2003.

\bibitem{Komissarov2006}
Komissarov, S. S.,``Magnetized Tori around Kerr Black Holes: Analytic Solutions with a Toroidal Magnetic Field,''\emph{Mon. Not. Roy. Astron. Soc.}, vol. 368, pp. 993-1000, 2006.

\bibitem{Tchekhovskoy2011}
Tchekhovskoy, Alexander and Narayan, Ramesh and McKinney, Jonathan C.,``Efficient Generation of Jets from Magnetically Arrested Accretion on a Rapidly Spinning Black Hole,''\emph{Mon. Not. Roy. Astron. Soc.}, vol. 418, pp. L79-L83, 2011.

\bibitem{Tejeda2012}
Tejeda, Emilio and Taylor, Paul A. and Miller, John C.,``An analytic toy model for relativistic accretion in Kerr spacetime,''\emph{Mon. Not. Roy. Astron. Soc.}, vol. 429, pp. 925, 2013.

\bibitem{Compere2017}
Comp\`ere, G. and Oliveri, R.,``Self-similar accretion in thin discs around near-extremal black holes,''\emph{Mon. Not. Roy. Astron. Soc.}, vol. 468, pp. 4351-4361, 2017.

\bibitem{Porth2017}
Porth, Oliver and Olivares, Hector and Mizuno, Yosuke and Younsi, Ziri and Rezzolla, Luciano and Moscibrodzka, Monika and Falcke, Heino and Kramer, Michael,``The black hole accretion code,''\emph{Comput. Astrophys. Cosmol.}, vol. 4, pp. 1, 2017.

\bibitem{Mitra2022}
Mitra, Samik and Maity, Debaprasad and Dihingia, Indu Kalpa and Das, Santabrata, ``Study of general relativistic magnetohydrodynamic accretion flow around black holes,''\emph{Mon. Not. Roy. Astron. Soc.}, vol. 516, pp. 5092-5109, 2022.

\bibitem{Hou2023}
Hou, Yehui and Zhang, Zhenyu and Guo, Minyong and Chen, Bin,``A new analytical model of magnetofluids surrounding rotating black holes,''\emph{JCAP}, vol.02, pp. 030, 2024.

\bibitem{Ginzburg1965}
V. L. Ginzburg and S. I. Syrovatskii, ``Developments in the Theory of Synchrotron Radiation and its Reabsorption,'' \emph{Ann. Rev. Astron. Astrophys.}, vol. 3, pp. 297每350, 1965.

\bibitem{Luminet1979}
Luminet, J. -P.,``Image of a spherical black hole with thin accretion disk,''\emph{Astron. Astrophys.}, vol. 75, pp. 228-235, 1979.

\bibitem{Bardeen1972}
Bardeen, James M. and Press, William H. and Teukolsky, Saul A,``Rotating black holes: Locally nonrotating frames, energy extraction, and scalar synchrotron radiation,''\emph{Astrophys. J.}, vol. 178, pp. 347. 1972.

\bibitem{Rybicki1979}
Rybicki, G. B., Lightman, A. P., ``Radiative Processes in Astrophysics,''\emph{ Wiley-VCH}, chapter 6 ※Synchrotron Radiation§, \S6.4-6.5.

\bibitem{14}
S. Chandrasekhar, \emph{The Mathematical Theory of Black Holes}, Oxford: Oxford University Press, 1983.

\bibitem{Cunningham1975}
Cunningham, C. T.,``The effects of redshifts and focusing on the spectrum of an accretion disk around a Kerr black hole,''\emph{Astrophys. J.}, vol. 202, pp. 788-802, 1975.

\bibitem{EHT2023}
Event Horizon Telescope Collaboration, ``First Sagittarius A* Event Horizon Telescope Results. V. Testing Astrophysical Models of the Galactic Center Black Hole,'' \emph{Astrophys. J. Lett.}, vol. 930, no. 2, p. L16, 2023.



\bibitem{Noble2009}
S. C. Noble, J. H. Krolik, and J. F. Hawley, ``Direct Calculation of the Radiative Efficiency of an Accretion Disk Around a Black Hole,'' \emph{Astrophys. J.}, vol. 692, no. 1, pp. 411每421, 2009.

\end{thebibliography}
\end{document}